\documentclass[a4paper,11pt]{article}
\usepackage{pos}
\usepackage{bm}

\title{Accidental symmetries in the scalar potential of the Standard Model extended with two Higgs triplets}
\ShortTitle{Accidental symmetries in the scalar potential of the 2HTM}

\author*[a,b,c]{Xin Wang}
\author[b,c]{Yilin Wang}
\author[b,c]{Shun Zhou}

\affiliation[a]{Department of Physics and State Key Laboratory of Nuclear Physics and Technology, Peking University,\\
Beijing 100871, China}

\affiliation[b]{
Institute of High Energy Physics, Chinese Academy of Sciences,\\
Beijing 100049, China}

\affiliation[c]{
School of Physical Sciences, University of Chinese Academy of Sciences,\\
Beijing 100049, China}

\emailAdd{x.wang@pku.edu.cn}
\emailAdd{wangyilin@ihep.ac.cn}
\emailAdd{zhoush@ihep.ac.cn}

\abstract{The two-Higgs-triplet model (2HTM) provides us with an attractive way to simultaneously account for tiny neutrino masses and the observed matter-antimatter asymmetry in our Universe. In this talk, we study the accidental symmetries of the scalar potential in the 2HTM using the bilinear-field formalism. Based on the group-theoretical arguments, we prove that the maximal symmetry group is ${\rm SO(4)}$. We carefully analyze all the subgroups of ${\rm SO(4)}$, and find that there are eight distinct kinds of accidental symmetries in the scalar potential of the 2HTM.}

\FullConference{%
  41st International Conference on High Energy physics - ICHEP2022\\
  6-13 July, 2022\\
  Bologna, Italy
}

\begin{document}
\maketitle

\section{Introduction}
It has been ten years since the discovery of the Higgs boson at the CERN Large Hadron Collider (LHC)~\cite{ATLAS:2012yve}. This extraordinary discovery has successfully proved the validity of the Higgs mechanism for the spontaneous gauge symmetry breaking, and completed the last piece of the puzzle of the Standard Model (SM). In addition, the subsequent experiments implemented by the ATLAS and CMS collaborations after the discovery also paint a clearer portrait of the Higgs boson. Despite the huge success, there are still some remaining puzzles related to the Higgs physics, e.g., nonzero neutrino masses, naturalness problem, matter-antimatter asymmetry, etc. These unsolved problems indicate some extended versions of the Higgs sector, which are expected to be examined over the next fifteen years~\cite{CMS:2022dwd}.

Among all the extensions of the Higgs sector, the two-Higgs-triplet model (2HTM), in which we extend the SM by introducing two triplet scalars with the same hypercharge $Y = -2$, is a very attractive one. On the one hand, the 2HTM can account for nonzero neutrino masses via the type-II seesaw mechanism~\cite{Konetschny:1977bn}, where the tiny neutrino masses can be attributed to the small vacuum expectation values of the Higgs triplets. On the other hand, a salient feature of the seesaw mechanisms is that they provide us with a natural way to explain the observed matter-antimatter asymmetry in our Universe via the thermal leptogenesis~\cite{Fukugita:1986hr}. However, if only one Higgs triplet is added to the SM, CP violation can not be produced in the out-of-equilibrium and lepton-number-violating decays of the Higgs triplet in the early Universe. Therefore, successful leptogenesis requires at least two Higgs triplets~\cite{Ma:1998dx}. 

As there are multiple Higgs fields in the 2HTM, the scalar potential becomes quite complex and leads to much richer phenomenology. To be specific, the scalar potential $V^{}_{\rm 2HTM}$ containing one Higgs doublet $H$ and two Higgs triplets ${\bm \phi}^{}_i = (\xi^1_i, \xi^2_i, \xi^3_i)^{\rm T}$ (for $i = 1, 2$) can be divided into the following three parts
\begin{eqnarray}
	V^{}_{\rm 2HTM} =  V^{}_{\rm H} + V^{}_{\phi} + V^{}_{{\rm H}\phi} \; . \label{eq:pot}
\end{eqnarray}
The pure-doublet potential $V^{}_{\rm H}$, the pure-triplet potential $V^{}_{\phi}$ and the doublet-triplet-mixing potential $V^{}_{{\rm H}\phi}$ in the above equation are respectively given by
\begin{equation}
	\begin{split}
	V^{}_{\rm H} =& -\mu^{2}_{\rm H}H^\dag_{} H + \lambda^{}_{\rm H} (H^\dag_{} H)^2_{} \; ,  \\
	V^{}_{\phi} =& \; m^2_{11}({\bm \phi}^*_1 \cdot {\bm \phi}^{}_1)  +m^2_{22}({\bm \phi}^*_2 \cdot {\bm \phi}^{}_2) + m^2_{12}({\bm \phi}^*_1 \cdot {\bm \phi}^{}_2) + m^{\ast 2}_{12}({\bm \phi}^{}_1 \cdot {\bm \phi}^*_2)  + \lambda^{}_1  ({\bm \phi}^*_1 \cdot {\bm \phi}^{}_1)^2_{} + \lambda^{}_2  ({\bm \phi}^*_2 \cdot {\bm \phi}^{}_2)^2_{}  \\
	&  + \lambda^{}_3  ({\bm \phi}^*_1 \cdot {\bm \phi}^{}_1) ({\bm \phi}^*_2 \cdot {\bm \phi}^{}_2) + \lambda^{}_4  ({\bm \phi}^*_1 \cdot {\bm \phi}^{}_2) ({\bm \phi}^{}_1 \cdot {\bm \phi}^*_2)  + \frac{\lambda^{}_5}{2} ({\bm \phi}^*_1 \cdot {\bm \phi}^{}_2)^2_{}  + \frac{\lambda^{\ast}_5}{2} ({\bm \phi}^{}_1 \cdot {\bm \phi}^*_2)^2_{}   \\
	& + ({\bm \phi}^*_1 \cdot {\bm \phi}^{}_1) \left[ \lambda^{}_6  ({\bm \phi}^*_1 \cdot {\bm \phi}^{}_2) + \lambda^{\ast}_6 ({\bm \phi}^{}_1 \cdot {\bm \phi}^*_2)\right] + ({\bm \phi}^*_2 \cdot {\bm \phi}^{}_2) \left[ \lambda^{}_7 ({\bm \phi}^*_1 \cdot {\bm \phi}^{}_2) + \lambda^{\ast}_7 ({\bm \phi}^{}_1 \cdot {\bm \phi}^*_2)\right] \\
	&  + \lambda^{}_8 ({\bm \phi}^*_1 \cdot {\bm \phi}^{\ast}_1)({\bm \phi}^{}_1 \cdot {\bm \phi}^{}_1) + \lambda^{}_9 ({\bm \phi}^{*}_2 \cdot {\bm \phi}^{\ast}_2)({\bm \phi}^{}_2 \cdot {\bm \phi}^{}_2) + \lambda^{}_{10} ({\bm \phi}^*_1 \cdot {\bm \phi}^\ast_2)({\bm \phi}^{}_1 \cdot {\bm \phi}^{}_2)   \\
	&  + \lambda^{}_{11} ({\bm \phi}^*_1 \cdot {\bm \phi}^\ast_1)({\bm \phi}^{}_2 \cdot {\bm \phi}^{}_2) + \lambda^\ast_{11} ({\bm \phi}^{}_1 \cdot {\bm \phi}^{}_1)({\bm \phi}^*_2 \cdot {\bm \phi}^\ast_2) + \lambda^{}_{12} ({\bm \phi}^*_1 \cdot {\bm \phi}^\ast_1)({\bm \phi}^{}_1 \cdot {\bm \phi}^{}_2)   \\
	& + \lambda^{\ast}_{12} ({\bm \phi}^{}_1 \cdot {\bm \phi}^{}_1)({\bm \phi}^\ast_1 \cdot {\bm \phi}^*_2) + \lambda^{}_{13} ({\bm \phi}^*_2 \cdot {\bm \phi}^\ast_2)({\bm \phi}^{}_1 \cdot {\bm \phi}^{}_2) + \lambda^{\ast}_{13} ({\bm \phi}^{}_2 \cdot {\bm \phi}^{}_2)({\bm \phi}^*_1  \cdot {\bm \phi}^\ast_2) \; ,  \\
	V^{}_{{\rm H}\phi} = &\; \lambda^{}_{14}(H^\dag_{}H)({\bm \phi}^*_1 \cdot {\bm \phi}^{}_1) + \lambda^{}_{15}(H^\dag_{}H)({\bm \phi}^*_2 \cdot {\bm \phi}^{}_2) + \lambda^{}_{16}(H^\dag_{}H) ({\bm \phi}^*_1 \cdot {\bm \phi}^{}_2) + \lambda^{\ast}_{16}(H^\dag_{}H) ({\bm \phi}^{}_1 \cdot {\bm \phi}^*_2)  \\
	& + \lambda^{}_{17}(H^\dag_{}{{\rm i} \bm \sigma} H)\cdot({\bm \phi}^\ast_1 \times {\bm \phi}^{}_1) + \lambda^{}_{18}(H^\dag_{}  {\rm i} {\bm \sigma} H)\cdot({\bm \phi}^\ast_2 \times {\bm \phi}^{}_2) + \lambda^{}_{19}(H^\dag_{}{\rm i} {\bm \sigma} H)\cdot({\bm \phi}^\ast_1 \times {\bm \phi}^{}_2)  \\
	& +  \lambda^{\ast}_{19}( H^\dag_{} {\rm i} {\bm \sigma} H)\cdot( {\bm \phi}^\ast_2 \times {\bm \phi}^{}_1) + (\mu^{}_1 H^{\rm T}_{}{\rm i}\sigma^{}_2{\bm \sigma}\cdot {\bm \phi}^{}_1 H + \mu^{}_2  H^{\rm T}_{}{\rm i}\sigma^{}_2{\bm \sigma}\cdot {\bm \phi}^{}_2 H + {\rm h.c.} )\; , 
	\label{eq:potpart}
	\end{split}
\end{equation}
where ${\bm \sigma}= (\sigma^1, \sigma^2, \sigma^3)^{\rm T}$ are the Pauli matrices with the transpose ``T'' acting only on the three-dimensional representation space. If the coupling constants in Eq.~(\ref{eq:potpart}) satisfy specific relations, the 2HTM may automatically possess some symmetries apart from the electroweak ${\rm SU(2)^{}_L} \otimes {\rm U(1)^{}_Y}$ gauge symmetry, which we call the accidental symmetries. Accidental symmetries can reduce the number of free parameters, and enhance the predictive power of the theory. Possible accidental symmetries in the two-Higgs-doublet model (2HDM) have been extensively investigated in the previous literature, see, e.g., Refs.~\cite{Ivanov:2007de,Battye:2011jj,Pilaftsis:2011ed}. In this work, we for the first time carry out a systematic study of all the accidental symmetries in the scalar potential of the 2HTM.

\section{The bilinear-field formalism}
\begin{figure}[t!]
	\centering		\includegraphics[width=0.75\textwidth]{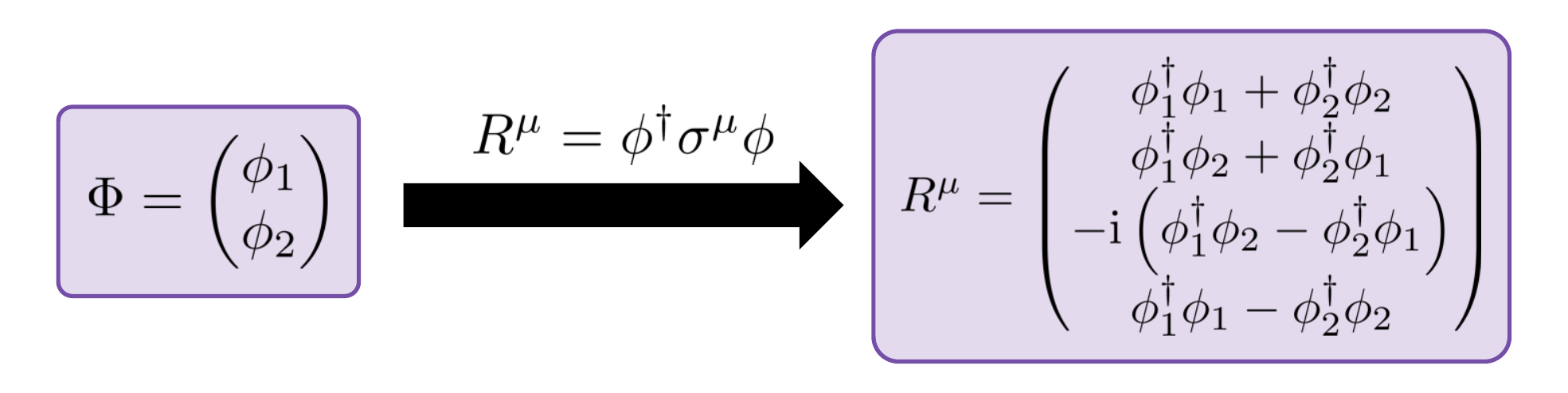} 	
	\caption{The bilinear transformation in the 2HDM.}
	\label{fig:2HDM} 
	\vspace{0cm}
\end{figure}

To start with, we briefly introduce the bilinear-field formalism, which is an effective way to study accidental symmetries in multi-Higgs models. Let us first take the 2HDM as an example~\cite{Ivanov:2005hg}. Suppose we have two Higgs doublets $\phi^{}_1$ and $\phi^{}_2$, which transform in the same way under the ${\rm SU(2)^{}_L}$ group. $\phi^{}_1$ and $\phi^{}_2$ form an ${\rm SU(2)}$ doublet $\Phi = (\phi^{}_1, \phi^{}_2)^{\rm T}_{}$. Inserting the Pauli matrices $\sigma^\mu_{} = (\sigma^0_{}, \bm{\sigma})$ (with $\sigma^0_{}$ being the two-dimensional identity matrix) between $\Phi^\dag_{}$ and $\Phi$, we can obtain a four-dimensional vector $R^{\mu}_{}$ (cf. Fig.~\ref{fig:2HDM}). It is easy to verify $R^\mu_{}$ is actually a vector in the Minkowski space. Hence the bilinear formalism reveals the geometrical properties of the 2HDM.

Now let us turn to the 2HTM case. Different from the 2HDM, both $\bm{\phi}$ and $\bm{\phi}^*_{}$ in the 2HTM transform in the same way under the ${\rm SU(2)^{}_L}$ group, motivating us to construct a four-dimensional multiplet $\Phi = ({\bm \phi}^{}_1,{\bm \phi}^{}_2,{\bm \phi}^{*}_1,{\bm \phi}^{*}_2)^{\rm T}_{}$. In fact, the four-component form of $\Phi$ is necessary to rewrite terms like $({\bm \phi}^*_i \cdot {\bm \phi}^{\ast}_j)({\bm \phi}^{}_k \cdot {\bm \phi}^{}_l)$ (for $i, j, k, l = 1, 2$) in Eq.~(\ref{eq:potpart}), and is also helpful to simultaneously investigate Higgs family symmetries and generalized CP symmetries in the scalar potential. Moreover, it is straightforward to check that $\Phi$ satisfies the Majorana condition $\Phi = {\rm C}\Phi^\ast_{}$ with ${\rm C} = \sigma^1_{} \otimes \sigma^{0}_{} \otimes {\bf I}^{}_{3\times3}$ being the charge conjugation matrix~\cite{Battye:2011jj,Pilaftsis:2011ed}. Similar to the 2HDM case, we can construct a vector $R^\mu$ in the bilinear-field space as $R^\mu_{} = \Phi^\dag_{} \Sigma^\mu_{} \Phi$ with $\Sigma^\mu_{} = \Sigma^\mu_{\alpha\beta} \sigma^\alpha_{} \otimes \sigma^\beta_{}$ (for $\alpha,\beta = 0,1,2,3$). $R^\mu_{}$ should keep invariant under the charge conjugation of $\Phi$, leading to the constraints $(\Sigma^\mu_{})^{\rm T}_{} = {\rm C}^{-1}_{} \Sigma^\mu_{} {\rm C}$. As a consequence, we arrive at ten nonzero $\Sigma^\mu_{}$ which can be written as
\begin{equation}
	\begin{split}
	&\Sigma^{0}_{} = +\dfrac{1}{2} \sigma^0_{} \otimes \sigma^0_{} ,  \;
	\Sigma^1_{} = -\dfrac{1}{2} \sigma^2_{} \otimes \sigma^3_{}  ,    \;
	\Sigma^2_{} = -\dfrac{1}{2} \sigma^1_{} \otimes \sigma^0_{}  ,    \;
	\Sigma^3_{} = +\dfrac{1}{2} \sigma^2_{} \otimes \sigma^1_{}  ,   \;
	 \Sigma^4_{} = -\dfrac{1}{2} \sigma^1_{} \otimes \sigma^3_{}  \; , \\
	&\Sigma^5_{} = +\dfrac{1}{2} \sigma^2_{} \otimes \sigma^0_{} ,\; 
	 \Sigma^6_{} = +\dfrac{1}{2} \sigma^1_{} \otimes \sigma^1_{}, \; 
	\Sigma^7_{} = +\dfrac{1}{2} \sigma^0_{} \otimes \sigma^1_{} , \;  \Sigma^8_{} = -\dfrac{1}{2} \sigma^3_{} \otimes \sigma^2_{} , \; 
	\Sigma^9_{} = +\dfrac{1}{2} \sigma^0_{} \otimes \sigma^3_{} \; .
	\label{eq:Sigmaform}
	\end{split}
\end{equation}
Substituting these nonzero $\Sigma^\mu_{}$ into $R^\mu_{}$, we can convert $\Phi$ into a ten-dimensional vector $R^\mu_{}$, the explicit form of which can be found in Ref.~\cite{Wang:2021ksj}.

Now that we have constructed the vector $R^\mu_{}$, the pure-triplet potential $V^{}_{\phi}$ then can be recast into the quadratic form
\begin{eqnarray}
	V^{}_{\phi} = \frac{1}{2} M^{}_\mu R^\mu + \frac{1}{4}  L^{}_{\mu\nu} R^\mu_{} R^\nu_{} \; ,
	\label{eq:pot2}
\end{eqnarray}
where both $M^{}_\mu$ and $L^{}_{\mu\nu}$ are coefficient matrices defined in Ref.~\cite{Wang:2021ksj}. 

\begin{figure}[t!]
	\centering		\includegraphics[width=0.92\textwidth]{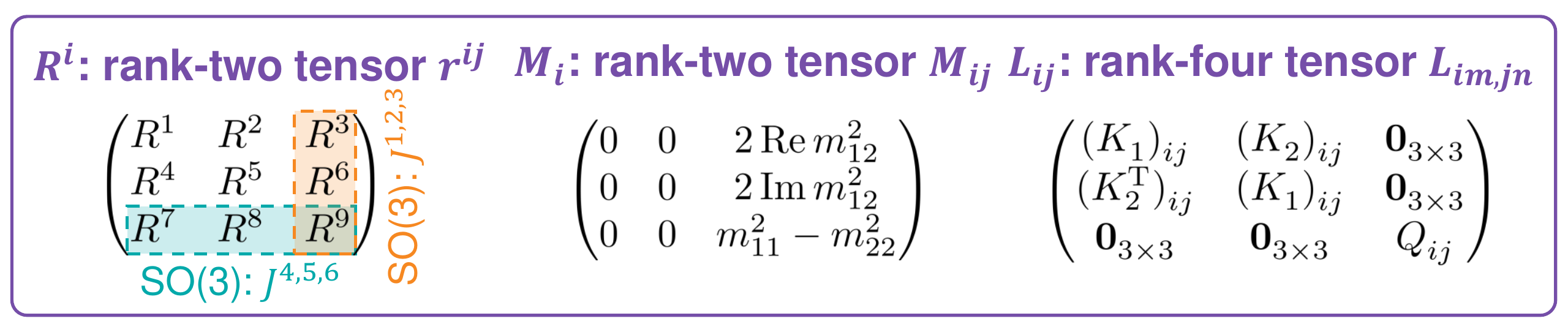} 	
	\caption{The vector $R^i_{}$ together with the coefficient matrices $M^{}_i$ and $L^{}_{ij}$ can be recast into three tensors $r^{ij}_{}$, $M^{}_{ij}$ and $L^{}_{im,jn}$.}
	\label{fig:tensor} 
	\vspace{0cm}
\end{figure}

\section{Accidental symmetries in the 2HTM}
With the help of the bilinear-field formalism, we are going to investigate the accidental symmetries in the 2HTM potential. The basic strategy is as follows. First, we determine the maximal symmetry group of the 2HTM. Then we carefully analyze the subgroups of the maximal symmetry group and find out all the continuous symmetries as well as discrete $Z^{}_2$ symmetries that the pure-triplet scalar potential $V^{}_\phi$ can accommodate. Combining them together we can arrive at the accidental symmetries of $V^{}_\phi$. Finally, we take the doublet-triplet-mixing potential into consideration and obtain the complete category of accidental symmetries in the 2HTM.

In order to determine the maximal symmetry group of the 2HTM, we should first require the kinetic terms of $\Phi$ to keep invariant under the symmetry transformation on $\Phi$, which leads to the unitary group ${\rm U(4)}$. The ${\rm U(4)}$ group will be further constrained by the Majorana conditions, i.e., the generators $J^a_{}$ (for $a = 1, 2, 3, \cdots, 16$) of the ${\rm U}(4)$ group should satisfy ${\rm C}^{-1}_{} J^{a}_{} {\rm C} = - (J_{}^{a})^\ast_{}$, which reduce the number of generators to six, namely,
\begin{equation}
	\begin{split}
	J_{}^{1} &= \frac{1}{2}\sigma^3_{} \otimes \sigma^3_{} \; , \quad J^{2}_{} = \frac{1}{2}\sigma^3_{} \otimes \sigma^1_{} \; , \quad J^{3}_{} = \frac{1}{2}\sigma^0_{} \otimes \sigma^2_{} \; ,  \\
	J_{}^{4} &= \frac{1}{2}\sigma^3_{} \otimes \sigma^0_{} \; , \quad J_{}^{5} = \frac{1}{2}\sigma^1_{} \otimes \sigma^2_{} \; , \quad J_{}^{6} = \frac{1}{2}\sigma^2_{} \otimes \sigma^2_{} \; .
	\end{split}
\end{equation}
It is not difficult to examine that the above generators satisfy the following Lie algebra
\begin{eqnarray}
	[J^{i}_{},J^{j}_{}] = {\rm i}\epsilon^{ijk}_{}J^{k}_{} \; , \quad [J^{i+3}_{},J^{j+3}_{}] = {\rm i}\epsilon^{ijk}_{}J^{k+3}_{} \; , \quad [J^{i}_{},J^{j+3}_{}] = 0  \; , \quad ({\rm for}~ i,j,k = 1,2,3)
	\label{eq:liealg}
\end{eqnarray}
with $\epsilon^{ijk}_{}$ being the three-dimensional Levi-Civita symbol. Hence one can immediately find that the maximal symmetry group in the $\Phi$-space is isomorphic to ${\rm SU}(2) \otimes {\rm SU}(2)$. Now we come back to the $R^\mu_{}$ space. The zero-component $R^0_{} = {\bm \phi}^*_1 \cdot {\bm \phi}^{}_1 + {\bm \phi}^*_2 \cdot {\bm \phi}^{}_2$ remains unchanged under the unitary transformations on $\Phi$, so we only focus on the ``spatial'' components $R^{i}_{}$. As there is a ``two-to-one correspondence'' between the $\Phi$- and $R^i_{}$-space, the maximal symmetry group in the $R^i_{}$-space turns out to be ${\rm SO}(4) \simeq [{\rm SU}(2) \times  {\rm SU}(2)]/Z^{}_2$.

Next we search for the possible accidental symmetries in the pure-triplet potential $V^{}_\phi$. As has been mentioned before, accidental symmetries in the 2HTM are the consequences of certain relations among the coupling constants. In order to find out these relations, we rearrange the components of $R^i_{}$ into a rank-two tensor $r^{ij}_{}$, whose elements are defined as $r^{ij}_{} = R^{3i+j-3}$ (for $i,j = 1,2,3$). Correspondingly, the spatial components of $M^{}_\mu$ and $L^{}_{\mu\nu}$ can also be recast into two tensors $M^{}_{ij}$ and $L^{}_{im,jn}$, as illustrated in Fig.~\ref{fig:tensor}. It is easy to identify that each column (or row) of $r^{ij}_{}$ can be regarded as the fundamental representation of the group ${\rm SO(3)}^i_{}$ [or ${\rm SO(3)}^j_{}$] generated by $\{J^1_{},J^2_{},J^3_{}\}$ (or $\{J^4_{},J^5_{},J^6_{}\}$), which allows us to implement the similar method in Ref.~\cite{Ivanov:2007de} to figure out all the accidental symmetries in the pure-triplet potential of the 2HTM.

Let us take the maximal symmetry group ${\rm SO(4)}$ for instance. Such a symmetry can be realized by requiring $K^{}_1$ and $Q$ to be proportional to the identity matrix ${\bf I}^{}_{3\times3}$, and all the other coupling matrices to be zero. As a result, we arrive at
\begin{equation}
	\begin{split}
		& m^2_{11}  = m^2_{22} \; , \quad
		m^2_{12}  = 0  \; , \quad
		\lambda^{}_1 = \lambda^{}_2  \; , \quad 
		\lambda^{}_3 = 2\lambda^{}_1 - 2\lambda^{}_8  \; , \\ 
		& \lambda^{}_4 = \lambda^{}_{10} = 2\lambda^{}_8 = 2 \lambda^{}_9 \; , \quad
		\lambda^{}_5 = \lambda^{}_{6} = \lambda^{}_7 = \lambda^{}_{11} = \lambda^{}_{12} =\lambda^{}_{13} = 0 \; .
	\end{split}
\end{equation}
Keeping the above relations in mind, we can express the ${\rm SO}(4)$-invariant potential $V^{}_{\phi,{\rm SO(4)}}$ as
\begin{equation}
	\begin{split}
		V^{}_{\phi,\,{\rm SO(4)}} =
		& \, m^2_{11}({\bm \phi}^*_1 \cdot {\bm \phi}^{}_1 + {\bm \phi}^*_2 \cdot {\bm \phi}^{}_2) + \lambda^{}_1  ({\bm \phi}^*_1 \cdot {\bm \phi}^{}_1+{\bm \phi}^*_2 \cdot {\bm \phi}^{}_2)^2_{} \\
		& + 2\lambda^{}_8 \left[ ({\bm \phi}^*_1 \cdot {\bm \phi}^{}_2)({\bm \phi}^{}_1 \cdot {\bm \phi}^*_2)-({\bm \phi}^*_1 \cdot {\bm \phi}^{}_1)({\bm \phi}^*_2 \cdot {\bm \phi}^{}_2) \right] \\
		& + \lambda^{}_8 \left[ ({\bm \phi}^*_1 \cdot {\bm \phi}^*_1)({\bm \phi}^{}_1 \cdot {\bm \phi}^{}_1)+ 2({\bm \phi}^*_1 \cdot {\bm \phi}^*_2)({\bm \phi}^{}_1 \cdot {\bm \phi}^{}_2)+({\bm \phi}^*_2 \cdot {\bm \phi}^*_2)({\bm \phi}^{}_2 \cdot {\bm \phi}^{}_2) \right] \; ,
	\end{split}
\end{equation}
where only three independent parameters $m^{}_{11}$, $\lambda^{}_1$ and $\lambda^{}_8$ are left. Similarly, one can analyze other continuous subgroups of ${\rm SO(4)}$ and find out all the continuous symmetries in $V^{}_\phi$.

Apart from the continuous symmetries, discrete $Z^{}_2$ symmetries may also exist in $V^{}_\phi$. This point can be easily understood since if $R^\mu$ and $R^\nu$ in Eq.~(\ref{eq:pot2}) simultaneously change their signs, the quadratic term of $V^{}_\phi$ will remain unchanged. We notice that $r^{ij}_{}$ could exhibit three different patterns if we act a $Z^{}_2$ transformation on $\Phi$, namely,
\begin{eqnarray}
	{\bf (a)}~\left(\begin{matrix}
		+ & + & - \\
		+ & + & - \\
		- & - & + \\
	\end{matrix}\right) \; ,
	\quad
	{\bf (b)}~\left(\begin{matrix}
		- & - & - \\
		+ & + & + \\
		- & - & - \\
	\end{matrix}\right) \; ,
	\quad
	{\bf (c)}~\left(\begin{matrix}
		+ & - & - \\
		+ & - & - \\
		+ & - & - \\
	\end{matrix}\right) \; .
	\label{eq:Rpattern}
\end{eqnarray}
Pattern {\bf(a)} means the elements in one row and one column change their signs; Pattern {\bf(b)} refers to the case where the elements in two rows flip their signs; Pattern {\bf (c)} corresponds to the scenario where the signs of the elements in two columns are reversed. The accidental symmetries should be the combination of all possible continuous symmetries and $Z^{}_2$ symmetries. For one continuous symmetry, we should identify whether it has already covered all the $Z^{}_2$ symmetries in the $R^i_{}$-space. If not, we need to include the $Z^{}_2$ symmetries that have been ignored in the previous analysis.

Finally we take the doublet-triplet-mixing terms into consideration. There are three different types of doublet-triplet-mixing terms, i.e., $(H^\dag_{}H)({\bm \phi}^*_{i} \cdot {\bm \phi}^{}_{j} )$, $(H^\dag_{}{\rm i}{\bm \sigma} H)\cdot({\bm \phi}^\ast_i \times {\bm \phi}^{}_j)$ and $H^{\rm T}_{}{\rm i}\sigma^{}_2{\bm \sigma}\cdot {\bm \phi}^{}_{i} H$. The inclusion of $(H^\dag_{}H)({\bm \phi}^*_{i} \cdot {\bm \phi}^{}_{j} )$ will not influence the accidental symmetries in the scalar potential, as they behave similarly as the bilinear terms ${\bm \phi}^*_{i} \cdot {\bm \phi}^{}_{j} $ under the transformations of ${\bm \phi}^{}_i$. If $(H^\dag_{}{\rm i}{\bm \sigma} H)\cdot({\bm \phi}^\ast_i \times {\bm \phi}^{}_j)$ terms are included, the maximal symmetry group will be reduced to ${\rm O(3)}^i_{} \otimes O(2)^j_{}$. The trilinear terms $H^{\rm T}_{}{\rm i}\sigma^{}_2{\bm \sigma}\cdot {\bm \phi}^{}_{i} H$ could violate all the accidental symmetries except the ${\rm SO}(2)^j_{}$ and $Z^{}_2$ symmetries. Therefore, the trilinear terms can be viewed as soft symmetry-breaking terms which are phenomenologically important in explaining tiny neutrino masses and avoiding undesired Goldstone bosons after the spontaneous symmetry breaking. After a careful analysis of all the accidental symmetries, we find that there are in total eight distinct types of accidental symmetries for the 2HTM potential. Please refer to Table~2 in Ref.~\cite{Wang:2021ksj} for the detailed classification.

\section{Summary}
We explore the accidental symmetries in the scalar potential of the 2HTM. We have proved that the maximal symmetry group in the 2HTM is SO(4), and demonstrated that there are eight different kinds of accidental symmetries in the scalar potential. These symmetries are useful for us to construct more predictive models with less parameters, investigate vacuum stability conditions and vacuum solutions, and study the topological structures of the 2HTM.

\vspace{0.2cm}

\emph{This work was supported in part by the National Natural Science Foundation of China under grant No. 11835013.}

\end{document}